\documentclass[journal]{IEEEtran}
\IEEEoverridecommandlockouts

\usepackage{cite}
\usepackage{graphics} 
\usepackage{epsfig} 
\usepackage{mathptmx} 
\usepackage{times} 
\usepackage{amsmath} 
\usepackage{amssymb} 
\usepackage{amsthm} 
\usepackage{dblfloatfix} 
\usepackage{comment}
\usepackage{mathtools}
\usepackage{tabularx,colortbl} 
\usepackage{multicol}
\usepackage{xcolor}

\usepackage[section]{placeins}
\usepackage{cuted}

\usepackage{tabularray}
\usepackage{float}
\usepackage[shortlabels]{enumitem}
\usepackage{placeins}
\DeclareMathAlphabet{\mathcal}{OMS}{cmsy}{m}{n}

\theoremstyle{remark}

\theoremstyle{definition}

\usepackage[utf8]{inputenc}
\usepackage[english]{babel}

\usepackage{multirow}
\usepackage{tabularx}
\usepackage{adjustbox}
\usepackage{array}   

\newcommand{\PreserveBackslash}[1]{\let\temp=\\#1\let\\=\temp}
\newcolumntype{C}[1]{>{\PreserveBackslash\centering}p{#1}}
\newcolumntype{R}[1]{>{\PreserveBackslash\raggedleft}p{#1}}
\newcolumntype{L}[1]{>{\PreserveBackslash\raggedright}p{#1}}

\begin{document}

\title{Discovery and Characterization of Cross-Area and Intra-Area SSOs Sensitive to Delay in Droop Control of Grid-Forming Converters}

\author{Lilan~Karunaratne,~\IEEEmembership{Student Member,~IEEE,}
        Nilanjan~Ray~Chaudhuri,~\IEEEmembership{Senior Member,~IEEE,}
        Amirthagunaraj~Yogarathnam,~\IEEEmembership{Member,~IEEE},
        and~Meng~Yue,~\IEEEmembership{Member,~IEEE}\vspace{-20pt} 
\thanks{This work is supported under Agreement 37532 by the Advanced Grid Modeling (AGM) Program of the Office of Electricity, the Department of Energy (DOE).}
\thanks{L. Karunaratne and N. R. Chaudhuri are with the School of Electrical Engineering and Computer
Science, Penn State University, University Park, PA, USA e-mail: lvk5363@psu.edu, nuc88@psu.edu.} 
\thanks{A. Yogarathnam and M. Yue are with the Interdisciplinary Science Department, Brookhaven National Laboratory, Upton, NY, USA e-mail: ayogarath@bnl.gov, yuemeng@bnl.gov.}
}

\maketitle
\begin{abstract}
Subsynchronous oscillations (SSOs) involving grid-forming converters (GFCs) are in a less familiar territory of power system dynamics. This letter reports a new phenomenon namely \textit{cross-area} SSOs in grids with 100\% droop-controlled GFC-based renewable penetration, which was discovered during our study on evaluating the adequacy of quasistationary phasor calculus (QPC) and space phasor calculus (SPC)-based models in capturing SSOs. 
We present frequency-domain characterization of such oscillatory modes in addition to intra-area SSOs in grids involving GFCs and study the impact of a delay in power-frequency droop feedback loop in regards to their stability. Electromagnetic transient (EMT) simulations validate our findings.

\end{abstract}


\section{Introduction}
Several incidents of subsynchronous oscillations (SSOs) involving grid-following converters (GFLCs) have been witnessed in bulk power systems in different parts of the world. Based on the root cause analysis, this phenomena can be divided into three categories - (a) series capacitor SSO \cite{IBRSSOTF}, (b) weak grid SSO \cite{IBRSSOTF}, and (c) inter-inverter-based resource (IBR) SSO \cite{Lingling-22-InterIBRosc}. Although papers including \cite{IBRSSOTF,Lingling-22-InterIBRosc} have explored SSO phenomena involving GFLCs, SSO involving grid-forming converters (GFCs) was only recently reported in \cite{interarea_osc} and in our work \cite{lilan_pesgm24}. 

Inter-IBR SSO reported in \cite{Lingling-22-InterIBRosc} can be categorized as \textit{intra-area} oscillations, i.e., IBRs in the same area oscillate against each other. On the other hand, \cite{interarea_osc} reported \textit{inter-area} oscillations under $100\%$ IBR penetration, where IBRs in different areas oscillate against each other. In essence, the oscillation groupings are characteristically similar to intra- and inter-area oscillations involving synchronous generators (SGs) \cite{kundur}.  

In this letter, we report the discovery of a different type of oscillatory mode in the IEEE $2$-area system \cite{kundur} with SGs replaced by droop-controlled GFCs. In our case, unlike typical inter-area oscillations, the IBR groups are formed across the areas and each group oscillates against the other (see Fig.~\ref{fig:11_bus_system}). Therefore, these oscillations can neither be categorized as intra- nor inter-area oscillations. We call this \textit{cross-area} SSO. 

In addition to the cross-area mode, we also observe intra-area SSOs in the IEEE 2-area system with 50\% and 75\% IBR penetrations. Frequency-domain characterization of such SSOs is performed and the impact of delay in the power-frequency droop control of GFCs towards the stability of these modes is assessed using space phasor calculus (SPC)-based models. The observations were validated using electromagnetic transient (EMT) models. We show that the quasistationary phasor calculus (QPC)-based models cannot capture the SSO modes.

\section{Modeling in SPC and QPC Frameworks}
Both SPC- and QPC-based models are built upon space phasors (also called space-vectors) denoted by $\bar{x}(t)$. 
The time-varying phasor $\bar{x}_{bb}(t)$ is calculated by the frequency shift operation on $\bar{x}(t)$ leading to $\bar{x}_{bb}(t) = x_d(t) + jx_q(t) = \bar{x}(t)e^{-j\rho(t)}$, where $\frac{\mathrm{d} \rho (t) }{\mathrm{d} t} = \omega (t)$. The operator $\bar{\Upsilon} : \mathcal{M} \rightarrow \mathcal{D}$ is defined as $\bar{x}_{bb}(t) = \bar{\Upsilon}(\textbf{x}(t))$, where $\mathcal{M}$ is a vector space representing the set of balanced three-phase signals. It can be shown that 
$\bar{\Upsilon}(\textbf{x}(t)) = [1~j~0] \mathcal{P}(t)\textbf{x}(t)$, where $\mathcal{P}(t)$ 
is the Park's transformation matrix. 
As opposed to SPC, QPC-based models consider certain approximations in modeling some of the components as summarized in Table~\ref{tab:modeling_of_components}. Dynamic models of SGs and GFCs are developed in respective rotating $d$-$q$ frames, while the transmission network and loads are represented in a synchronously rotating $D$-$Q$ frame. For more details, readers are referred to the previous works \cite{lilan_pesgm24,lilan_2024modeling} of the authors. 
Nonetheless, for the sake of completeness, the key aspects of GFC modeling and controls are concisely presented in the following subsection.

\begin{table}[h!]
\scriptsize
\caption{Summary of component modeling in different frameworks}
\label{tab:modeling_of_components}
\centering
\begin{tabular}{|L{2.6cm}| *{2}{C{2.5cm}|}}
    \hline
    \multirow{2}{*}{\textbf{Component}} &
        \multicolumn{2}{c|}{\textbf{Modeling framework}} \\
        \cline{2-3}
        & \textbf{QPC} & \textbf{SPC}\\
    \hline
    SG stator transients & neglected & dynamically modeled \\
    \hline
    Transmission network & Y-bus & dynamically modeled\\
    \hline
    Loads & static load model & dynamic load model \\
    \hline
    GFC & \multicolumn{2}{c|}{identically modeled} \\
    \hline
\end{tabular}
\end{table}

\subsection{Modeling of GFC and its controls}
\begin{figure}[h!]
    \centerline{
    \includegraphics[width=0.42\textwidth]{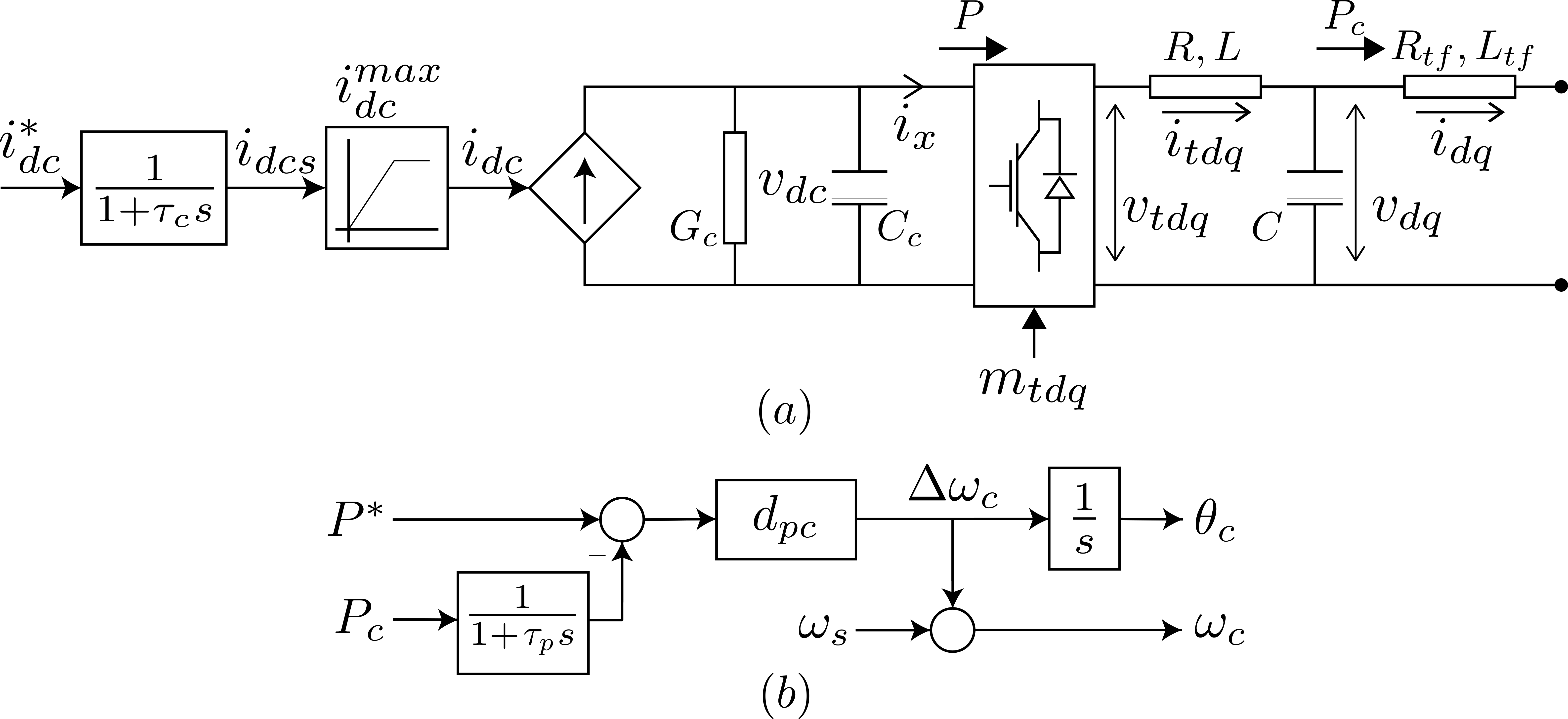 }}
    \vspace{-10pt}
    \caption{(a) Circuit model of GFC with dc-side representing functional model of a renewable source, (b) power-frequency droop with delay $\tau_p$.}
    \label{fig:Circuit diagram of GFC}
    \vspace{-10pt}
\end{figure}

A generic circuit diagram of a GFC is illustrated in Fig.~\ref{fig:Circuit diagram of GFC}(a). The notations used are self-explanatory and adhere to conventional meanings. The ac-side of the GFC is modeled and controlled in SPC framework that employs a rotating $d$-$q$ frame whose angular frequency $\omega_c$ is determined by the power frequency droop mechanism depicted in Fig.~\ref{fig:Circuit diagram of GFC}(b).  

\begin{figure}[h!]
    \centerline{
    \includegraphics[width=0.422\textwidth]{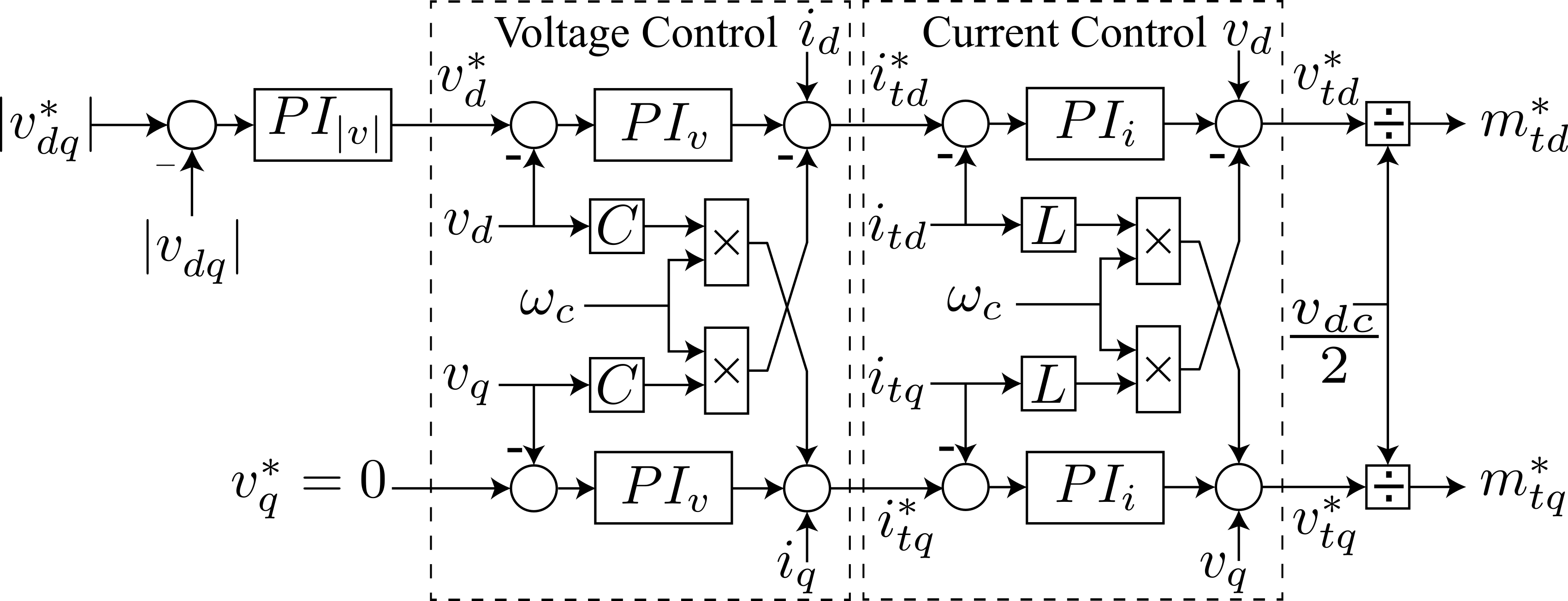}}
    \vspace{-5pt}
    \caption{Inner current and voltage control loops, and outer voltage control loop of GFC.}
    \label{fig:Inner loop current and voltage controls of GFC}
    \vspace{-5pt}
\end{figure}

Additionally, the standard inner current and voltage control loops, and outer voltage control loop are employed to regulate ac-side quantities as shown in Fig.~\ref{fig:Inner loop current and voltage controls of GFC}. 

\section{Case Studies}
\begin{figure}[h!]
    \centerline{
    \includegraphics[width=0.5\textwidth]{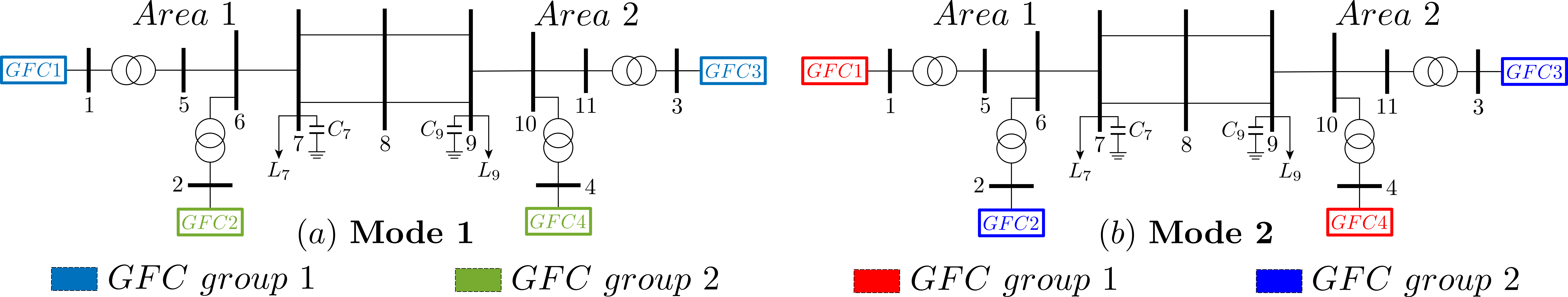}}
    \vspace{-5pt}
    \caption{Case 4: \textit{Cross-area} SSO modes forming distinct GFC groups in modified IEEE $2$-area test system with 100\% IBR penetration. }
    \label{fig:11_bus_system}
    \vspace{-0pt}
\end{figure}

\noindent \textit{A. \underline{Test systems and modeling frameworks}:} In this section, we delve into characterization of the SSO modes involving GFCs using 
case studies performed on the IEEE $2$-area $4$-machine test system \cite{kundur} modified by different IBR penetration levels after replacing some or all of the SGs by GFCs. Moreover, the impact of delay $\tau_p$ in power-frequency droop feedback loop (Fig.~\ref{fig:Circuit diagram of GFC}(b)) is studied to understand its influence on such modes. All test systems considered operate with approximately $400$~MW tie-line flow. Our analysis will concerntrate on four specific case studies: case ($1$): $50$\% IBR penetration replacing G$1$ and G$4$; case ($2$): $50$\% IBR penetration replacing G$1$ and G$2$; case ($3$): $75$\% IBR penetration replacing G$1$, G$2$, and G$3$; and case ($4$): $100$\% IBR penetration (see, Fig.~\ref{fig:11_bus_system}). However, due to space constraints, we will discuss case ($4$) in details and findings from other cases are summarized.

To that end, QPC and SPC-based models of the test systems are developed in Matlab/Simulink \cite{simulink}.  EMT models of the same systems are built in EMTDC/PSCAD \cite{pscad}, where averaged models of the converters are considered along with distributed parameter representation of transmission lines, and (dynamic) constant impedance models of the loads.

\begin{figure}[h!]
    \centerline{
    \includegraphics[width=0.45\textwidth]{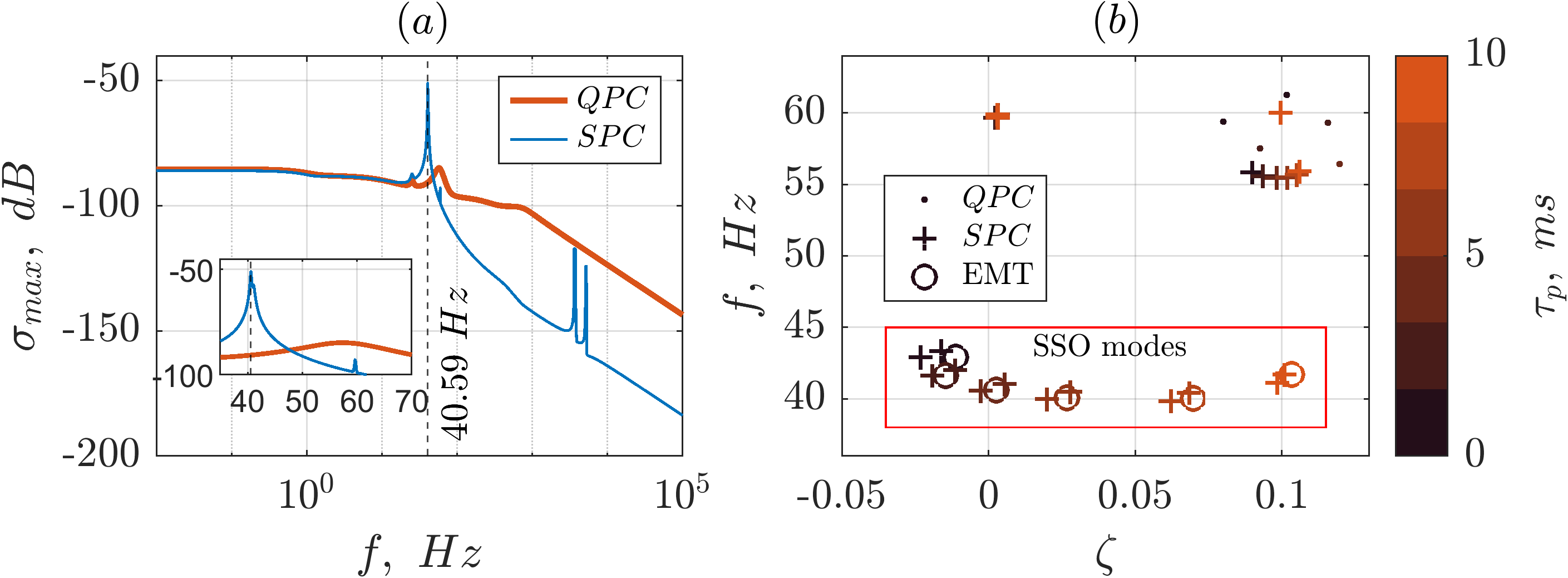}}
    \vspace{-5pt}
    \caption{Case ($4$): (a) Comparison of maximum singular values of QPC and SPC models when $\tau_p = 2$~ms, (b) the loci of the SSO modes as the delay $\tau_p$ varies.}
    \label{fig:com_plot}
    \vspace{0pt}
\end{figure}

\noindent \textit{B. \underline{Modeling adequacy- SPC vs QPC}:} We perform the modeling adequacy study including frequency-domain analysis as proposed in \cite{lilan_2024modeling} on the linearized QPC and SPC-based models. For all cases, the QPC-based models cannot capture the SSO modes and are therefore deemed inadequate, whereas the SPC-based models can capture these modes. As an example, the maximum singular value (SV) plots of the $4$-input-$12$-output linearized models based on QPC and SPC with $\tau_p = 2$~ms for case ($4$) are illustrated in Fig.~\ref{fig:com_plot}(a), where a signal modulating active power references to the GFCs are inputs, and the dc-link voltages and transformer currents of the GFCs are outputs. 

The SPC-based model exhibits a sharp peak in gain around \textcolor{black}{$40.59$}~Hz representing an unstable mode with $\zeta = -0.27$\% (Mode $2$ in Fig.~\ref{fig:11_bus_system}) as shown in Fig.~\ref{fig:com_plot}(b), while the QPC-based model lacks this characteristic. Figure~\ref{fig:com_plot}(b) indicates another SSO mode of \textcolor{black}{$41.06$}~Hz with $\zeta = 0.55$\% (Mode $1$ in Fig.~\ref{fig:11_bus_system}), which is stable and does not appear in the SV plot in Fig.~\ref{fig:com_plot}(a). Both the SSO modes are absent in the QPC model (Fig.~\ref{fig:com_plot}(b)). This observation is consistent with QPC and SPC-based models of rest of the cases, and indicates inadequacy of QPC models. 

\begin{figure}[h!]
    \centerline{
    \includegraphics[width=0.48\textwidth]{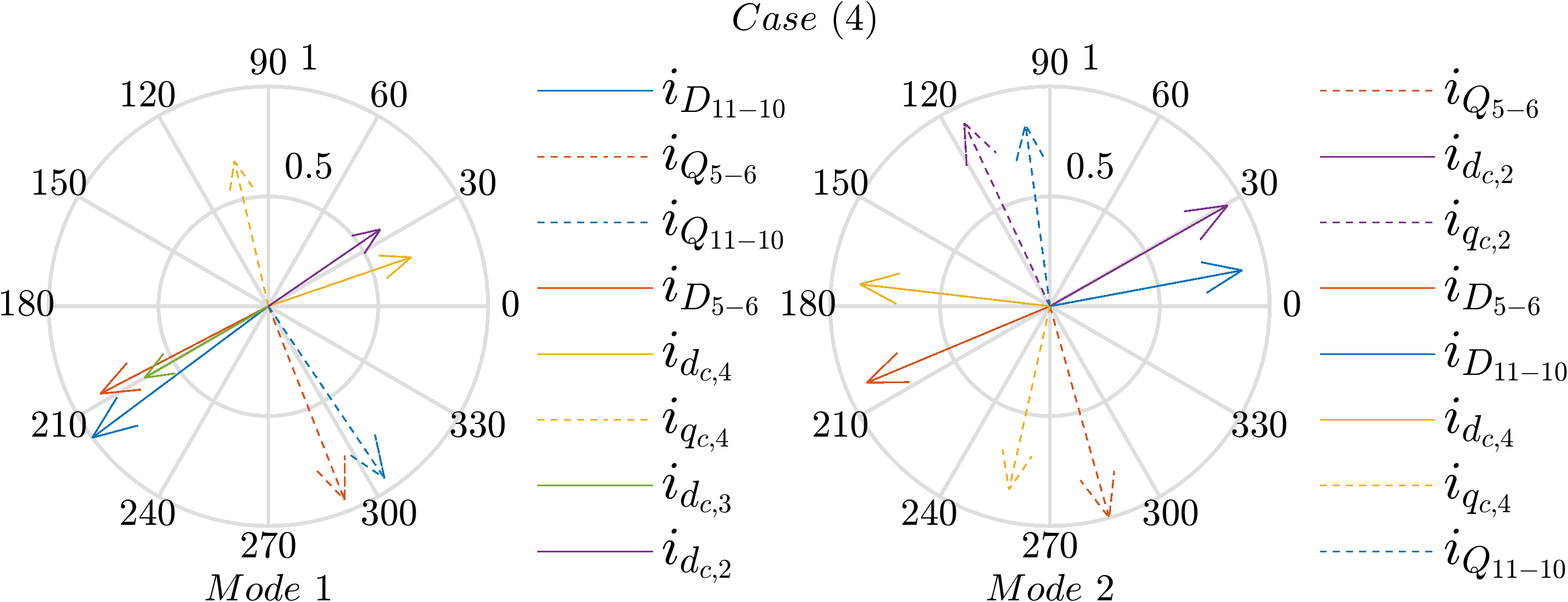}}
    \vspace{-5pt}
    \caption{Case (4): Compass plots of normalized participation factor magnitudes and modeshape angles of the dominant states contributing to the \textit{cross-area} SSO modes for $\tau_p = 2$~ms.}
    \label{fig:compass_plot_2}
    \vspace{0pt}
\end{figure}

\noindent \textit{C. \underline{Characterization of SSO modes from SPC-based models}:} 
The SSO mode characterization is performed through frequency-domain analysis of all four cases. The results shown in this section assume \textcolor{black}{$\tau_p = 2$~ms for SPC-based models}.\\ 
(I) \underline{Cross-area mode}: First, we focus on the characterization of the most interesting SSO modes, which we call \textit{cross-area} modes. 
Figure~\ref{fig:compass_plot_2} illustrates the compass plots of normalized participation factor magnitudes and modeshape angles of the dominant states contributing to these modes (higlighted in Fig.~\ref{fig:com_plot}(b)). The most dominant states associated with the SSO modes are terminal and transformer currents of GFCs and current states of transmission line segments (i.e., $i_{DQ_{5-6}}$, $i_{DQ_{11-10}}$) that connect some of the GFCs to the network.

Unlike traditional \textit{inter-area} modes, where groups of SGs or IBRs in one area oscillate against another group in an electrically-distant area \cite{kundur,interarea_osc}, we observe a different characteristic in these modes. \textit{Here, two GFCs in different areas group together and oscillate against the remaining two GFCs across those areas, which we refer to as \underline{\textit{cross-area}} modes. }

The modeshape angles relating to Mode $2$ indicate that GFC$1$ and GFC$4$ as a group oscillate against GFC$2$ and GFC$3$, see Fig.~\ref{fig:11_bus_system}(b). Conversely, the modeshape angles for Mode $1$ suggest that GFC$1$ and GFC$3$ oscillate against GFC$2$ and GFC$4$ as shown in Fig.~\ref{fig:11_bus_system}(a). Regardless of how the GFCs are grouped, an intriguing commonality is that the groupings consistently involve GFCs from two different areas -- a phenomenon not observed before.

\begin{figure}[h!]
    \centerline{
    \includegraphics[width=0.45\textwidth]{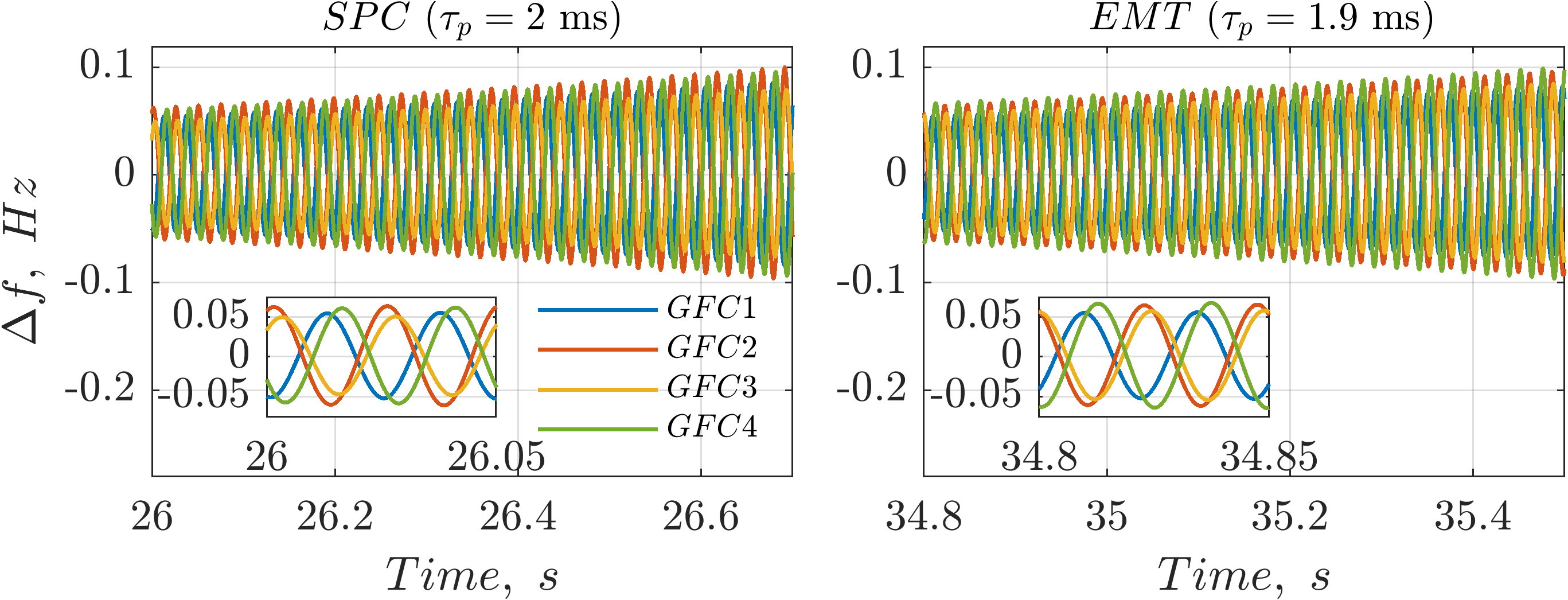}}
    \vspace{-5pt}
    \caption{Case ($4$): Comparison of frequency dynamics of GFCs obtained from SPC and EMT simulations showing cross-area Mode 2.}
    \label{fig:freq_plot_2}
    \vspace{0pt}
\end{figure}

Based on the singular value plot from the SPC model in Fig.~\ref{fig:com_plot}(a), Mode $2$ in this case is expected to dominate the dynamic behavior of the system. This fact is further verified by the time-domain responses of frequency dynamics from SPC and EMT simulations shown in Fig.~\ref{fig:freq_plot_2}, which also validate the GFC groupings in the cross-area Mode $2$. For EMT simulation $\tau_p = 1.9$ ms was used, which corroborated with the SPC case with $\tau_p = 2$ ms. Prony analysis on the EMT model's responses reveals that the frequency of oscillation is \textcolor{black}{$40.69$}~Hz with \textcolor{black}{$\zeta = -0.24$\%}, which is pretty close to what we observe for Mode $2$ of the SPC model.

\begin{figure}[h!]
    \centerline{
    \includegraphics[width=0.48\textwidth]{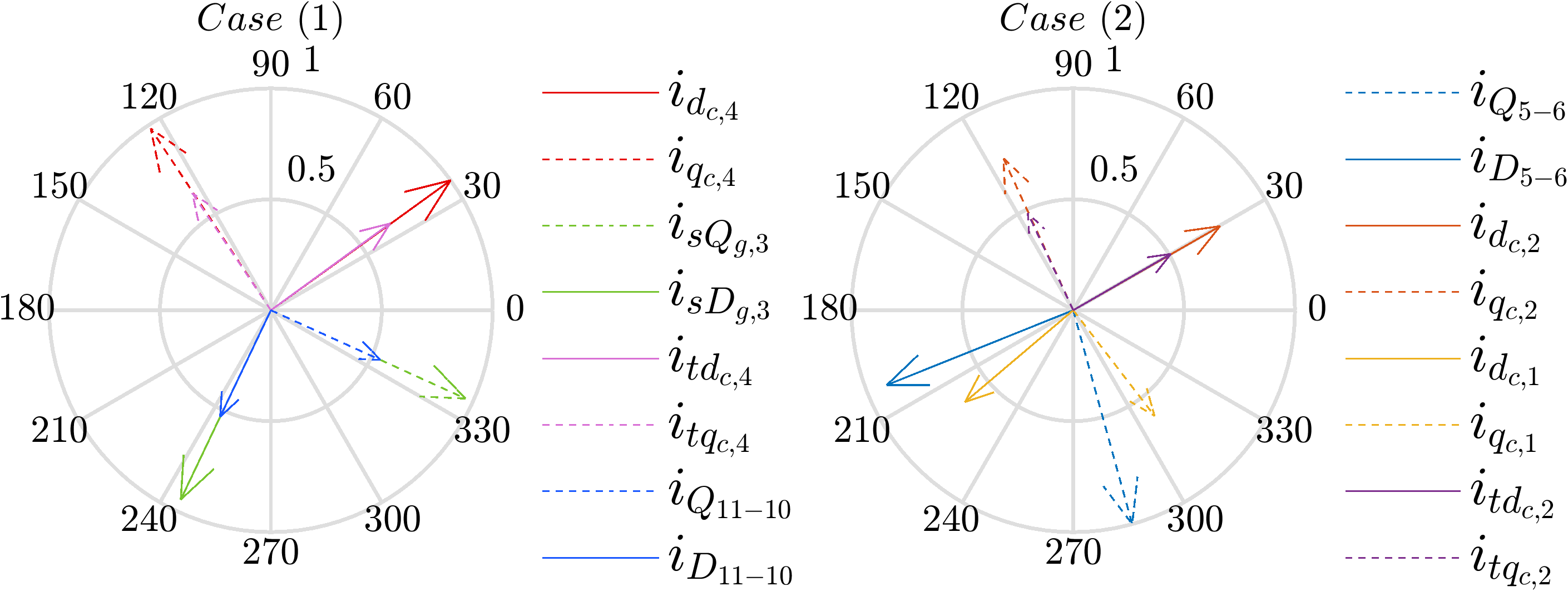}}
    \vspace{-5pt}
    \caption{Case ($1$) and ($2$): Compass plots of normalized participation factor magnitudes and modeshape angles of the dominant states contributing to the \textit{intra-area} SSO modes for $\tau_p = 2$~ms.}
    \label{fig:compass_plot_1}
    \vspace{0pt}
\end{figure}

(II) \underline{Intra-area mode}:
For case ($1$) and case ($2$), the SSO modes observed can be categorized as \textit{intra-area} modes, since in both cases, the machines involved belong to the same area (see Fig.~\ref{fig:compass_plot_1}). An important observation is that for case ($1$) when the GFC interacts with a SG within the same area, the resulting SSO mode remains stable. Conversely, for \textcolor{black}{case ($2$)} an interaction between two GFCs within the same area may lead to an unstable SSO mode for certain values of $\tau_p$ as shown in Table~\ref{tab:unstable_sso_modes}.  For case ($3$), two separate intra-area modes, one among GFC$1$ and GFC$2$ and the other among G$3$ and GFC4 are observed, where the former is unstable for certain $\tau_p$ values (Table~\ref{tab:unstable_sso_modes}) and the latter is always stable (not shown in Table~\ref{tab:unstable_sso_modes}). 

\noindent \textit{D. \underline{Impact of delay ($\tau_p$) on stability of SSO modes:}} We found that the SSO modes are sensitive to the delay $\tau_p$ of the power feedback loop of the droop controller shown in Fig.~\ref{fig:Circuit diagram of GFC}(b). Figure~\ref{fig:com_plot}(b) indicates the movement of SSO modes in case ($4$) as $\tau_p$ is varied from $0$ to $10$~ms. The results from EMT simulations are closely aligned with those of SPC model although only one of the modes is clearly observable in the EMT responses. Impact of delay for SSO modes of interest in cases ($2$) and ($3$) are shown in Table~\ref{tab:unstable_sso_modes}. The modes become stable for all cases when delay becomes $3$ ms or higher.

\begin{table}[ht]
    \centering
    \caption{SSO mode(s) of interest}
    \label{tab:unstable_sso_modes}
    \scriptsize
    \vspace{-5pt}
    \begin{adjustbox}{width=0.49\textwidth}
    \begin{tabular}{ccccc|cccccc}
    \hline
    
    \multicolumn{5}{|l|}{\multirow{1}{*}{\textbf{\begin{tabular}[c]{@{}l@{}} \textbf{$Delay$ ($\tau_p$)$,~ms$} \end{tabular}}}} 
    & \multicolumn{1}{c|}{\textbf{\begin{tabular}[c]{@{}l@{}} $0$ \end{tabular}}}
    & \multicolumn{1}{c|}{\textbf{\begin{tabular}[c]{@{}l@{}} $1$ \end{tabular}}}
    & \multicolumn{1}{c|}{\textbf{\begin{tabular}[c]{@{}l@{}} $2$ \end{tabular}}}
    & \multicolumn{1}{c|}{\textbf{\begin{tabular}[c]{@{}l@{}} $3$ \end{tabular}}}
    & \multicolumn{1}{c|}{\textbf{\begin{tabular}[c]{@{}l@{}} $5$ \end{tabular}}}
    & \multicolumn{1}{c|}{\textbf{\begin{tabular}[c]{@{}l@{}} $10$ \end{tabular}}}
    \\ \hline
    
    \multicolumn{1}{|l|}{\multirow{4}{*}{\textbf{\begin{tabular}[c]{@{}l@{}}\rotatebox[origin=c]{90}{Case ($2$)} \\ \end{tabular}}}} 
    & \multicolumn{1}{l}{\multirow{4}{*}{\textbf{\begin{tabular}[c]{@{}l@{}}\rotatebox[origin=c]{90}{\textit{intra-area}} \\ \end{tabular}}}} 
    & \multicolumn{1}{l|}{\multirow{4}{*}{\textbf{\begin{tabular}[c]{@{}l@{}}\hspace{-8pt}\rotatebox[origin=c]{90}{\textit{mode}} \\ 
    \end{tabular}}}} 
    & \multicolumn{1}{l|}{\multirow{2}{*}{\textbf{\begin{tabular}[c]{@{}l@{}}\rotatebox[origin=c]{90}{SPC} \end{tabular}}}} 
    & \multicolumn{1}{l|}{\textbf{\begin{tabular}[c]{@{}l@{}} $f, Hz$ \end{tabular}}}
    & \multicolumn{1}{c|}{{\begin{tabular}[c]{@{}l@{}}\rowcolor{lightgray} $43.18$ \end{tabular}}}
    & \multicolumn{1}{c|}{{\begin{tabular}[c]{@{}l@{}}\rowcolor{lightgray} $41.88$ \end{tabular}}}
    & \multicolumn{1}{c|}{{\begin{tabular}[c]{@{}l@{}}\rowcolor{lightgray} $40.88$ \end{tabular}}}
    & \multicolumn{1}{c|}{{\begin{tabular}[c]{@{}l@{}}\rowcolor{lightgray} $40.30$ \end{tabular}}}
    & \multicolumn{1}{c|}{{\begin{tabular}[c]{@{}l@{}}\rowcolor{lightgray} $40.20$ \end{tabular}}}
    & \multicolumn{1}{c|}{{\begin{tabular}[c]{@{}l@{}}\rowcolor{lightgray} $41.46$ \end{tabular}}} 
    \\ \cline{5-11}
    \multicolumn{1}{|l|}{} & \multicolumn{1}{l}{} & \multicolumn{1}{l|}{} &\multicolumn{1}{l|}{} & \multicolumn{1}{l|}{\textbf{\begin{tabular}[c]{@{}l@{}} $\zeta, \%$ \end{tabular}}}
    & \multicolumn{1}{c|}{{\begin{tabular}[c]{@{}l@{}} -$2.02$ \end{tabular}}}
    & \multicolumn{1}{c|}{{\begin{tabular}[c]{@{}l@{}} -$1.58$ \end{tabular}}}
    & \multicolumn{1}{c|}{{\begin{tabular}[c]{@{}l@{}} $0.08$ \end{tabular}}}
    & \multicolumn{1}{c|}{{\begin{tabular}[c]{@{}l@{}} $2.32$ \end{tabular}}}
    & \multicolumn{1}{c|}{{\begin{tabular}[c]{@{}l@{}} $6.46$ \end{tabular}}}
    & \multicolumn{1}{c|}{{\begin{tabular}[c]{@{}l@{}} $9.89$ \end{tabular}}} 
    \\ \cline{4-11}
    \multicolumn{1}{|l|}{} & \multicolumn{1}{l}{} & \multicolumn{1}{l|}{} &\multicolumn{1}{l|}{\multirow{2}{*}{\textbf{\begin{tabular}[c]{@{}l@{}}\rotatebox[origin=c]{90}{EMT} \end{tabular}}}} 
    & \multicolumn{1}{l|}{\textbf{\begin{tabular}[c]{@{}l@{}} $f, Hz$ \end{tabular}}}
    & \multicolumn{1}{c|}{{\begin{tabular}[c]{@{}l@{}}\rowcolor{lightgray} $43.13$ \end{tabular}}}
    & \multicolumn{1}{c|}{{\begin{tabular}[c]{@{}l@{}}\rowcolor{lightgray} $41.90$ \end{tabular}}}
    & \multicolumn{1}{c|}{{\begin{tabular}[c]{@{}l@{}}\rowcolor{lightgray} $40.94$ \end{tabular}}}
    & \multicolumn{1}{c|}{{\begin{tabular}[c]{@{}l@{}}\rowcolor{lightgray} $40.35$ \end{tabular}}}
    & \multicolumn{1}{c|}{{\begin{tabular}[c]{@{}l@{}}\rowcolor{lightgray} $40.17$ \end{tabular}}}
    & \multicolumn{1}{c|}{{\begin{tabular}[c]{@{}l@{}}\rowcolor{lightgray} $41.38$ \end{tabular}}}
    \\ \cline{5-11}
    \multicolumn{1}{|l|}{} & \multicolumn{1}{l}{} & \multicolumn{1}{l|}{} & \multicolumn{1}{l|}{} & \multicolumn{1}{l|}{\textbf{\begin{tabular}[c]{@{}l@{}} $\zeta,\%$ \end{tabular}}}
    & \multicolumn{1}{c|}{{\begin{tabular}[c]{@{}l@{}} -$1.33$ \end{tabular}}}
    & \multicolumn{1}{c|}{{\begin{tabular}[c]{@{}l@{}} -$0.86$ \end{tabular}}}
    & \multicolumn{1}{c|}{{\begin{tabular}[c]{@{}l@{}} $0.30$ \end{tabular}}}
    & \multicolumn{1}{c|}{{\begin{tabular}[c]{@{}l@{}} $2.54$ \end{tabular}}}
    & \multicolumn{1}{c|}{{\begin{tabular}[c]{@{}l@{}} $6.60$ \end{tabular}}}
    & \multicolumn{1}{c|}{{\begin{tabular}[c]{@{}l@{}} $10.10$ \end{tabular}}}
    \\ \hline \hline
    
    \multicolumn{1}{|l|}{\multirow{4}{*}{\textbf{\begin{tabular}[c]{@{}l@{}}\rotatebox[origin=c]{90}{Case ($3$)} \\ \end{tabular}}}} 
    & \multicolumn{1}{l}{\multirow{4}{*}{\textbf{\begin{tabular}[c]{@{}l@{}}\rotatebox[origin=c]{90}{\textit{intra-area}} \\ \end{tabular}}}} 
    & \multicolumn{1}{l|}{\multirow{4}{*}{\textbf{\begin{tabular}[c]{@{}l@{}}\hspace{-8pt}\rotatebox[origin=c]{90}{\textit{mode}} \\ 
    \end{tabular}}}} 
    & \multicolumn{1}{l|}{\multirow{2}{*}{\textbf{\begin{tabular}[c]{@{}l@{}}\rotatebox[origin=c]{90}{SPC} \end{tabular}}}} 
    & \multicolumn{1}{l|}{\textbf{\begin{tabular}[c]{@{}l@{}} $f, Hz$ \end{tabular}}}
    & \multicolumn{1}{c|}{{\begin{tabular}[c]{@{}l@{}}\rowcolor{lightgray} $43.13$ \end{tabular}}}
    & \multicolumn{1}{c|}{{\begin{tabular}[c]{@{}l@{}}\rowcolor{lightgray} $41.82$ \end{tabular}}}
    & \multicolumn{1}{c|}{{\begin{tabular}[c]{@{}l@{}}\rowcolor{lightgray} $40.82$ \end{tabular}}}
    & \multicolumn{1}{c|}{{\begin{tabular}[c]{@{}l@{}}\rowcolor{lightgray} $40.23$ \end{tabular}}}
    & \multicolumn{1}{c|}{{\begin{tabular}[c]{@{}l@{}}\rowcolor{lightgray} $40.11$ \end{tabular}}}
    & \multicolumn{1}{c|}{{\begin{tabular}[c]{@{}l@{}}\rowcolor{lightgray} $41.37$ \end{tabular}}} 
    \\ \cline{5-11}
    \multicolumn{1}{|l|}{} & \multicolumn{1}{l}{} & \multicolumn{1}{l|}{} & \multicolumn{1}{l|}{} & \multicolumn{1}{l|}{\textbf{\begin{tabular}[c]{@{}l@{}} $\zeta, \%$ \end{tabular}}}
    & \multicolumn{1}{c|}{{\begin{tabular}[c]{@{}l@{}} -$2.11$ \end{tabular}}}
    & \multicolumn{1}{c|}{{\begin{tabular}[c]{@{}l@{}} -$1.69$ \end{tabular}}}
    & \multicolumn{1}{c|}{{\begin{tabular}[c]{@{}l@{}} -$0.03$ \end{tabular}}}
    & \multicolumn{1}{c|}{{\begin{tabular}[c]{@{}l@{}} $2.22$ \end{tabular}}}
    & \multicolumn{1}{c|}{{\begin{tabular}[c]{@{}l@{}} $6.37$ \end{tabular}}}
    & \multicolumn{1}{c|}{{\begin{tabular}[c]{@{}l@{}} $9.85$ \end{tabular}}} 
    \\ \cline{4-11}
    \multicolumn{1}{|l|}{} & \multicolumn{1}{l}{} & \multicolumn{1}{l|}{} & \multicolumn{1}{l|}{\multirow{2}{*}{\textbf{\begin{tabular}[c]{@{}l@{}}\rotatebox[origin=c]{90}{EMT} \end{tabular}}}} 
    & \multicolumn{1}{l|}{\textbf{\begin{tabular}[c]{@{}l@{}} $f, Hz$ \end{tabular}}}
    & \multicolumn{1}{c|}{{\begin{tabular}[c]{@{}l@{}}\rowcolor{lightgray} $43.09$ \end{tabular}}}
    & \multicolumn{1}{c|}{{\begin{tabular}[c]{@{}l@{}}\rowcolor{lightgray} $41.87$ \end{tabular}}}
    & \multicolumn{1}{c|}{{\begin{tabular}[c]{@{}l@{}}\rowcolor{lightgray} $40.85$ \end{tabular}}}
    & \multicolumn{1}{c|}{{\begin{tabular}[c]{@{}l@{}}\rowcolor{lightgray} $40.27$ \end{tabular}}}
    & \multicolumn{1}{c|}{{\begin{tabular}[c]{@{}l@{}}\rowcolor{lightgray} $40.08$ \end{tabular}}}
    & \multicolumn{1}{c|}{{\begin{tabular}[c]{@{}l@{}}\rowcolor{lightgray} $41.29$ \end{tabular}}}
    \\ \cline{5-11}
    \multicolumn{1}{|l|}{} & \multicolumn{1}{l}{} & \multicolumn{1}{l|}{} &\multicolumn{1}{l|}{} & \multicolumn{1}{l|}{\textbf{\begin{tabular}[c]{@{}l@{}} $\zeta,\%$ \end{tabular}}}
    & \multicolumn{1}{c|}{{\begin{tabular}[c]{@{}l@{}} -$1.48$ \end{tabular}}}
    & \multicolumn{1}{c|}{{\begin{tabular}[c]{@{}l@{}} -$0.98$ \end{tabular}}}
    & \multicolumn{1}{c|}{{\begin{tabular}[c]{@{}l@{}} $0.22$ \end{tabular}}}
    & \multicolumn{1}{c|}{{\begin{tabular}[c]{@{}l@{}} $2.45$ \end{tabular}}}
    & \multicolumn{1}{c|}{{\begin{tabular}[c]{@{}l@{}} $6.50$ \end{tabular}}}
    & \multicolumn{1}{c|}{{\begin{tabular}[c]{@{}l@{}} $10.03$ \end{tabular}}}
    \\ \hline 
    
    \end{tabular}
    \end{adjustbox}
\end{table}

\section{Conclusion}
A detailed analysis was conducted using participation factors and mode shapes of the dominant states associated with SSO modes involving GFCs. This analysis revealed a new \textit{cross-area} oscillatory mode, where GFCs from different areas group together and oscillate against the remaining GFCs. Specific to the IEEE $2$-area system, the study further demonstrated that the SSO mode involving a GFC and SG from the same area remains stable. In contrast, interactions between two GFCs within the same area may result in an unstable SSO mode for this system. Moreover, the critical influence of a delay in the power-frequency droop feedback loop on the stability of SSO modes involving GFCs was examined. A delay of 3 ms was found to stabilize the unstable SSO modes in all the cases considered.

\bibliographystyle{IEEEtran}
\bibliography{reference.bib}

\begin{thebibliography}{1}
\providecommand{\url}[1]{#1}
\csname url@samestyle\endcsname
\providecommand{\newblock}{\relax}
\providecommand{\bibinfo}[2]{#2}
\providecommand{\BIBentrySTDinterwordspacing}{\spaceskip=0pt\relax}
\providecommand{\BIBentryALTinterwordstretchfactor}{4}
\providecommand{\BIBentryALTinterwordspacing}{\spaceskip=\fontdimen2\font plus
\BIBentryALTinterwordstretchfactor\fontdimen3\font minus \fontdimen4\font\relax}
\providecommand{\BIBforeignlanguage}[2]{{%
\expandafter\ifx\csname l@#1\endcsname\relax
\typeout{** WARNING: IEEEtran.bst: No hyphenation pattern has been}%
\typeout{** loaded for the language `#1'. Using the pattern for}%
\typeout{** the default language instead.}%
\else
\language=\csname l@#1\endcsname
\fi
#2}}
\providecommand{\BIBdecl}{\relax}
\BIBdecl

\bibitem{IBRSSOTF}
Y.~Cheng, L.~Fan, J.~Rose, S.-H. Huang, J.~Schmall, X.~Wang, X.~Xie, J.~Shair, J.~R. Ramamurthy, N.~Modi, C.~Li, C.~Wang, S.~Shah, B.~Pal, Z.~Miao, A.~Isaacs, J.~Mahseredjian, and J.~Zhou, ``Real-world subsynchronous oscillation events in power grids with high penetrations of inverter-based resources,'' \emph{IEEE Transactions on Power Systems}, vol.~38, no.~1, pp. 316--330, 2023.

\bibitem{Lingling-22-InterIBRosc}
L.~Fan, ``Inter-{IBR} oscillation modes,'' \emph{IEEE Transactions on Power Systems}, vol.~37, no.~1, pp. 824--827, 2022.

\bibitem{interarea_osc}
M.~Zhang, Z.~Miao, L.~Fan, and S.~Shah, ``Data-driven interarea oscillation analysis for a 100\% {IBR}-penetrated power grid,'' \emph{IEEE Open Access Journal of Power and Energy}, vol.~10, pp. 93--103, 2023.

\bibitem{lilan_pesgm24}
\BIBentryALTinterwordspacing
L.~Karunaratne, N.~R. Chaudhuri, A.~Yogarathnam, and M.~Yue, ``A case study on modeling adequacy of a grid with subsynchronous oscillations involving ibrs,'' 2024. [Online]. Available: \url{https://arxiv.org/abs/2407.09526}
\BIBentrySTDinterwordspacing

\bibitem{kundur}
P.~Kundur, \emph{Power System Stability and Control}.\hskip 1em plus 0.5em minus 0.4em\relax McGraw-Hill, 1994.

\bibitem{lilan_2024modeling}
\BIBentryALTinterwordspacing
L.~Karunaratne, N.~R. Chaudhuri, A.~Yogarathnam, and M.~Yue, ``An approach to evaluate modeling adequacy for small-signal stability analysis of {IBR}-related {SSO}s in multimachine systems,'' 2024 (under review). [Online]. Available: \url{https://doi.org/10.48550/arXiv.2403.07835}
\BIBentrySTDinterwordspacing

\bibitem{simulink}
\emph{{MATLAB version 9.10.0.2198249 (R2021a)}}, The Mathworks, Inc., Natick, Massachusetts, 2021.

\bibitem{pscad}
\emph{{PSCAD version 5.0.2.0(2023/04/14)}}, Manitoba HVDC Research Centre, Winnipeg, Canada, 2023.

\end{thebibliography}

\end{document}